\documentclass{rnaastex}

\begin{document}

\title{A Fruit of a Different Kind: 2015~BP$_{519}$ as an Outlier among the Extreme Trans-Neptunian Objects}

\correspondingauthor{Carlos~de~la~Fuente~Marcos}
\email{nbplanet@ucm.es}

\author[0000-0003-3894-8609]{Carlos~de~la~Fuente~Marcos}
\affiliation{Universidad Complutense de Madrid \\
             Ciudad Universitaria, E-28040 Madrid, Spain}

\author[0000-0002-5319-5716]{Ra\'ul~de~la~Fuente~Marcos}
\affiliation{AEGORA Research Group \\
             Facultad de Ciencias Matem\'aticas \\
             Universidad Complutense de Madrid \\
             Ciudad Universitaria, E-28040 Madrid, Spain}


\section{} 

The presence of one or more planets orbiting the Sun between the orbit of Neptune and the Oort Cloud has been investigated multiple times 
for many decades, but all the efforts to observe them have so far been unsuccessful. This subject found new significance in the wake of the 
discovery of 2012~VP$_{113}$ \citep{2014Natur.507..471T}, when a number of intriguing patterns were identified in the distributions of the 
orbital elements of the known extreme trans-Neptunian objects ---those with semimajor axis greater than 150~au and perihelion distance 
greater than 30~au--- or ETNOs \citep{2014MNRAS.443L..59D,2014Natur.507..471T}. The patterns were tentatively explained as caused by one 
\citep{2014Natur.507..471T} or more \citep{2014MNRAS.443L..59D} trans-Plutonian planets. Further evidence in the form of theoretical and 
numerical analyses was presented ---Planet Nine hypothesis--- by \citet{2016AJ....151...22B}. However, the causality in which a perturber 
induces a peculiar orbital architecture on the ETNOs has been dismissed by some researchers \citep{2017AJ....154...50S}, who claim that 
observational biases in the data are the true cause. The announcement of the discovery of 2015~BP$_{519}$ \citep{2018AJ....156...81B}, 
unofficially known as {\it Caju}, has been hailed by some as a consistent piece of robust evidence for the existence of a 
yet-to-be-discovered massive planet far beyond the trans-Neptunian belt. Here, we use the latest data available from 
\href{http://ssd.jpl.nasa.gov/sbdb.cgi}{JPL's Small-Body Database} \citep{2015IAUGA..2256293G} to show that 2015~BP$_{519}$ is a statistical 
outlier within the 29 known ETNOs; therefore, it cannot be used as a reference to further support the trends perhaps present for other ETNOs.

\begin{figure}[!ht]
\begin{center}
\includegraphics[scale=0.3,angle=0]{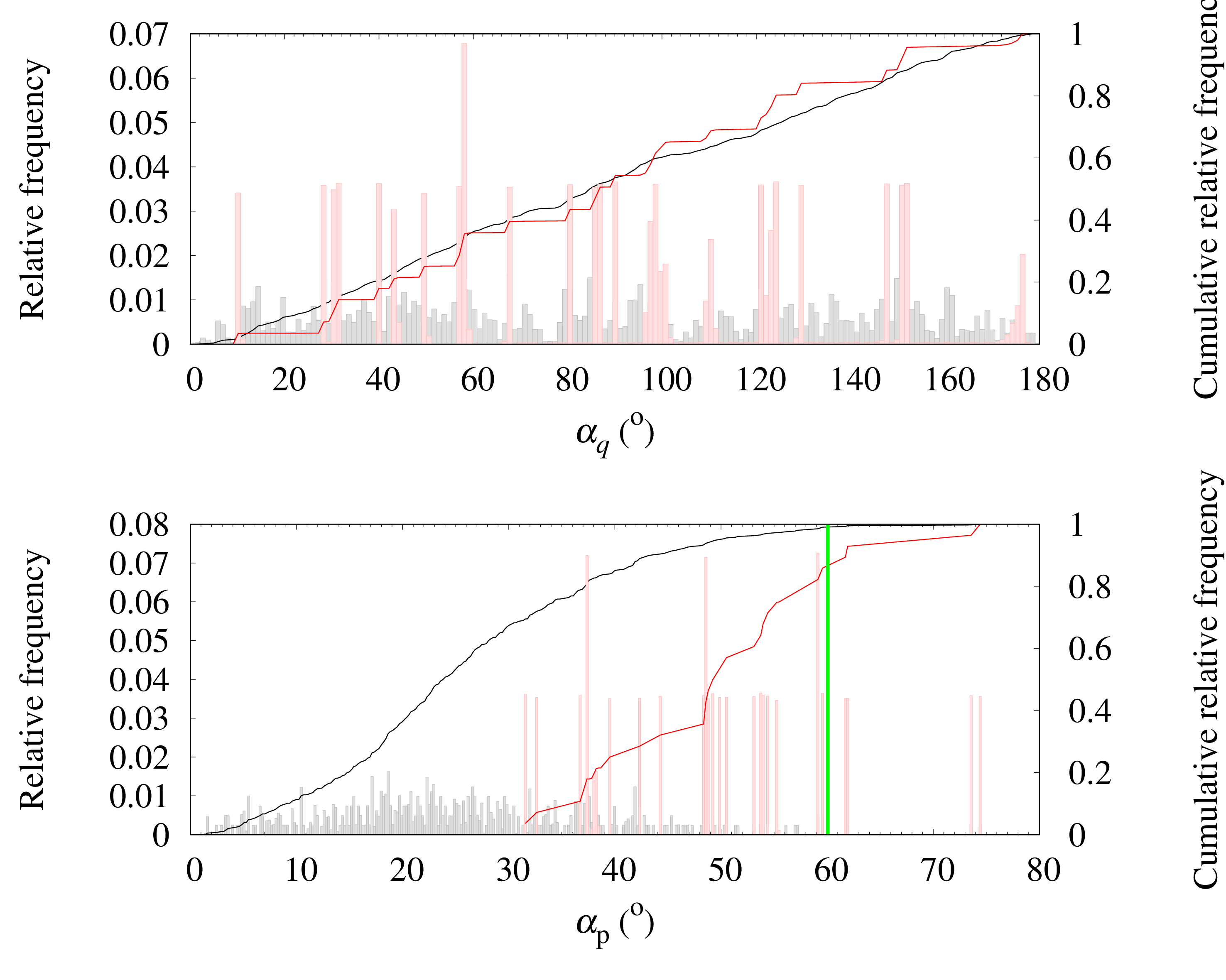}
\includegraphics[scale=0.35,angle=0]{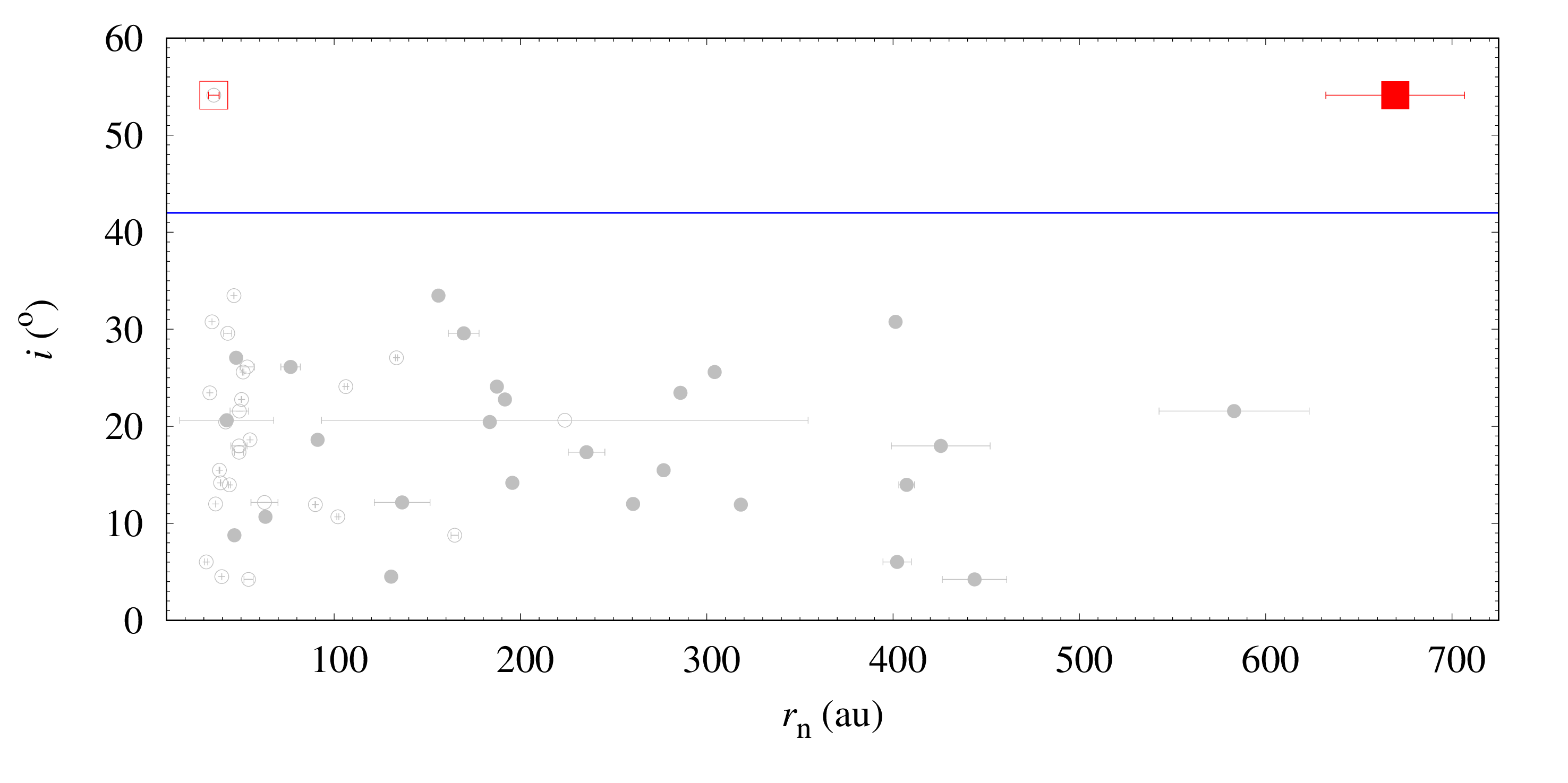}
\caption{Distributions of the angular separations between perihelia, $\alpha_q$ (top panel, the bin size is 1\fdg1), and orbital poles, 
         $\alpha_{\rm p}$ (middle panel, 0\fdg22), upper outlier limit in green. The black curve/gray bins are the result of the analysis of 
         4$\times$10$^{6}$ random pairs of ETNOs with synthetic orbits based on the mean values and dispersions of the orbital elements of 
         real ETNOs. The bin width has been computed using the Freedman-Diaconis rule \citep{FD81}: $2\ {\rm IQR}\ n^{-1/3}$, where $n$ is 
         the number of data points. In red/pink we show the distributions of pairs including 2015~BP$_{519}$. Distribution of nodal 
         distances (bottom panel, descending nodes in solid color, ascending empty), in red 2015~BP$_{519}$, upper outlier limit in blue.
\label{fig:1}}
\end{center}
\end{figure}

The orbit determination of 2015~BP$_{519}$ (epoch JD~2458200.5, 23-March-2018, solution date 31-May-2018) is based on 30 observations for a 
data-arc span of 1176 days and has semimajor axis, $a=430\pm24$~au, eccentricity, $e=0.918\pm0.005$, inclination, $i=54\fdg1173\pm0\fdg0003$,
longitude of the ascending node, $\Omega=135\fdg192\pm0\fdg005$, and argument of perihelion, $\omega=348\fdg39\pm0\fdg07$; 
\citet{2018AJ....156...81B} already mentioned the unusual value of $i$. The patterns present in ETNO orbital parameter space produce 
clustering in the location of perihelia and poles, and the study of the distributions of pole and perihelion separations (between each pair, 
see \citealt{2016MNRAS.462.1972D,2017Ap&SS.362..198D}) can help in identifying features inconsistent with the overall behavior. 
Figure~\ref{fig:1} shows such distributions (in gray) computed considering the errors in the values of the orbital parameters involved. 

Following \citet{2016MNRAS.462.1972D} and using the criteria discussed by \citet{1977eda..book.....T}, we searched for outliers in the 
distributions of pole and perihelion separations. Here, outliers are observations that fall below Q$_{1}-1.5$ IQR (lower limit) or above 
Q$_{3}+1.5$ IQR (upper limit), where Q$_{1}$ is the first quartile, Q$_{3}$ is the third quartile, and IQR is the interquartile range. The 
upper outlier limit in $i$ is $42\fdg0$ (in blue in Figure~\ref{fig:1}, bottom panel), but the inclination of 2015~BP$_{519}$ is above 
54\degr. If the other ETNOs reached their current orbital inclinations as a result of the same process (be it due to massive perturbers or 
any other), it is difficult to claim that whichever mechanisms are responsible for the current orbital organization of most of the ETNOs 
could also be behind the present-day orbit of 2015~BP$_{519}$. A different origin for the orbit of 2015~BP$_{519}$ is further supported by 
the fact that its orbital pole is well away from those of the others (pairs including 2015~BP$_{519}$ in pink in Figure~\ref{fig:1}). Having 
relatively well-aligned orbital poles is indicative of a fairly consistent direction of orbital angular momentum, which in turn suggests 
that the objects are subjected to the same type of background perturbation. Figure~\ref{fig:1}, middle panel, shows the upper outlier limit 
for the pole separations, $60\fdg0$ (in green); all the pairs with separations above this value include 2015~BP$_{519}$, i.e. the overall 
orientation in space of 2015~BP$_{519}$ is rather different from those of the other known ETNOs. In addition and following 
\citet{2017MNRAS.471L..61D}, the nodal distances of 2015~BP$_{519}$, in red in Figure~\ref{fig:1}, bottom panel, also suggest an outlier 
nature.

In this Note, we have provided evidence against 2015~BP$_{519}$ having followed the same dynamical pathway that placed the ETNOs where they 
are now. Asteroid 2015~BP$_{519}$ represents a case of extreme dynamical anticorrelation within the ETNO orbital realm; the objects 
discussed by \citet{2017MNRAS.467L..66D} and \citet{2017Ap&SS.362..198D} represent the opposite side, that of well-correlated pairs of 
ETNOs. 

\acknowledgments

We thank A.~I. G\'omez de Castro for providing access to computing facilities. This work was partially supported by the Spanish MINECO under 
grant ESP2015-68908-R. In preparation of this Note, we made use of the NASA Astrophysics Data System and the MPC data server.

\end{document}